\documentclass[%
reprint,
 twocolumn,
 showpacs,
 aps,
pra,
 longbibliography
]{revtex4-1}
\usepackage{amssymb, graphicx,subfigure}
\usepackage{enumerate,tensor}
\usepackage{amsmath,bm}
\usepackage{dcolumn}
\usepackage{color}
\usepackage{natbib} 
\usepackage[latin1]{inputenc}
\usepackage{subfigure}
\usepackage{textcomp}
\usepackage{booktabs}
\usepackage{makecell}
\usepackage{psfrag}
\usepackage{epsfig}

\usepackage[per-mode=symbol]{siunitx}   
\usepackage[version=3]{mhchem}
 
\DeclareSIUnit\intensity{\watt\per\centi\meter\squared}
\DeclareSIUnit\fieldstrength{\volt\per\centi\meter}
\DeclareSIUnit\kfieldstrength{k\volt\per\centi\meter}
\DeclareSIUnit\energy{cm^{-1}}
\DeclareSIUnit\pol{\AA^{3}}

\newcommand{\degree}{\ensuremath{^\circ}}%
\newcommand{\ie}{i.\,e.}%
\newcommand{\Ialign}{\textup{I}_{0}}%
\newcommand{\tensornote}[1]{\ensuremath{\underline{\underline{#1}}}}

\makeatletter

\newcommand{\ket}[1]{\left|#1\right\rangle}

\newcommand{\escalar}[2]{\left\langle #1|#2\right\rangle}

\newcommand{\expected}[1]{\left\langle #1\right\rangle}
\newcommand{\aarphys}{\affiliation{Department of Physics and Astronomy, Aarhus University, 8000
      Aarhus C, Denmark}}%
\newcommand{\uoftchem}{\affiliation{Chemical Physics Theory Group, Department of Chemistry,
and Center for Quantum Information and Quantum Control, University of Toronto,
Toronto, ON M5S 3H6, Canada}}%

\begin{document}

\title{Theoretical study of asymmetric superrotors: alignment and orientation}

\author{Juan J.\ Omiste}
\email{jomiste@chem.utoronto.ca}
\uoftchem 
\aarphys

\date{\today}
\begin{abstract} 

We report a theoretical study of the optical centrifuge acceleration of an asymmetric top molecule interacting with an electric static field by solving the time-dependent Schr\"odinger equation in the rigid rotor approximation.
 A detailed analysis of the mixing of the angular momentum in both the molecular and the laboratory fixed frames allow us to  deepen the understanding of the main features of the acceleration process, for instance, the effective angular frequency of the molecule at the end of the pulse. In addition, we prove numerically that the asymmetric superrotors rotate around one internal axis and that their dynamics is confined to the plane defined by the polarization axis of the laser, in agreement with experimental findings. Furthermore, we consider the orientation patterns induced by the dc field, showing the characteristics of their structure as a function of the strength of the static field and the initial configuration of the fields.

\end{abstract}
\maketitle
\section{Introduction}
\label{sec:introduction}

The control of molecular dynamics using external fields has reached an unprecedent degree of precision in the last decades~\cite{Lemeshko2013}. Specifically, the possibility of aligning and orienting molecular ensembles in space has significant implications in many areas, including the study of chemical reactions~\cite{brooks:science,brooks:jcp45,Loesch1994,Loesch1994a}, steric effects in collisions~\cite{Janssen1991} or the description of molecules using X-ray and electron diffraction~\cite{Nakajima2015}, among others.

One of the most prominent techniques to constrain the motion of molecules in space relies on the application of non-resonant nano- to femtosecond laser pulses, which allow to align a molecule along one axis of the laboratory~\cite{sakai:jcp110,poulsen:jcp121} or, furthermore, along the three axes using circular or elliptically polarized laser fields~\cite{larsen:phys_rev_lett_85_2470}. On the other hand, the orientation adds a direction to the alignment. The simplest method to orient an ensemble of polar molecules is the application of a static field~\cite{bulthuis:jpca101,loesch:jcp93,friedrich:zpd18}. The combination of both, a dc field and a non-resonant laser pulse allows for a large degree of orientation and alignment in the adiabatic limit for linear~\cite{friedrich:jcp111,Sakai2003,omiste:pra2012,Nielsen2012} and asymmetric molecules~\cite{kupper:jcp131,Omiste2011, Omiste2013, Hansen2013,Omiste2016_asymmetric_molecules,Thesing2017}, but it is extremely difficult to reach in general~\cite{Omiste2016_asymmetric_molecules, Thesing2017}. Many other experimental setups have been proposed to enhance the alignment of a molecular ensemble using external fields, such as the application of two laser pulses with different polarizations~\cite{Bisgaard2006,lee:phys_rev_lett_97,viftrup:pra79}, the application of single cycle pulses~\cite{Ortigoso2012} and single THz pulses~\cite{Damari2016,Zhang2017}, or a combination of THz and femtosecond laser~\cite{Kitano2011a,Zai2015} to efficiently achieve both orientation and alignment.

A different insight to the control of the molecular motion using non-resonant laser fields is achieved by means of the optical centrifuge, an IR linearly polarized laser pulse whose polarization axis rotates with linearly increasing angular velocity~\cite{Karczmarek1999,Villeneuve2000}. This type of laser fields are able to drive the molecular rotation, which follows the laser's polarization axis up to high angular velocities~\cite{Korobenko2014}, leading the molecules to superrotor states. These states are characterized not only by their fast rotation, but also by the constrain of their dynamics to the plane defined by the centrifuge. For highly rotating states, the centrifugal force may lead to the stretching of bonds~\cite{Korobenko2015a} or even to dissociation~\cite{Karczmarek1999}. Furthermore, it has been shown that an ensemble of superrotors is stable against collisions~\cite{Forrey2001,Tilford2004,Korobenko2014}, which allows to measure the spin-rotation coupling~\cite{Milner2014}, the interaction with a magnetic field~\cite{Floss2015,Milner2015b,Korobenko2015b} or may also cause magnetization in an ensemble of paramagnetic superrotors~\cite{Milner2017}.

Many properties of the superrotors can be understood classically, for instance the constraining of the molecular motion to the plane defined by the polarization axis of the laser induced by the centrifugal force~\cite{Karczmarek1999}. However, phenomena as the revivals in the alignment observed in the experiment~\cite{Korobenko2015a} demand a full quantum approach~\cite{Hartmann2012}. In this work, we tackle a full quantum mechanical description of the alignment and orientation of an asymmetric molecule prototype, SO$_2$, interacting with an optical centrifuge and a static electric field by solving the time-dependent Schr\"odinger equation (TDSE).

This paper is organized as follows: In Sec.~\ref{sec:system_and_numerical_methods} the system under study, its full Hamiltonian and the numerical techniques to solve the TDSE of the system are presented. In Sec.~\ref{sec:results} we discuss the numerical results for the optical centrifuge interacting with SO$_2$. Specifically, we describe in Sec.~\ref{sec:centrifugal} the centrifugal dynamics in terms of the mixing of the angular momentum along different directions. In Sec.~\ref{sec:alignment_polarization_plane}, we analyze the alignment of the SO$_2$ for several optical centrifuge parameters and in Sec.~\ref{sec:orienting_superrotors} we study the orientation induced by the electric static field. In Sec.~\ref{sec:conclusions}, we summarize the main conclusions of this work and the outlook for future projects. Finally, we include in the Appendices~\ref{sec:derivation_of_the_laser_term} and \ref{sec:coupling_wigner} the derivation of the Hamiltonian corresponding to the interaction with the fields and a summary of the Wigner D-matrix elements, respectively.

\section{The system and numerical methods}
\label{sec:system_and_numerical_methods}
\begin{figure}[t]
  \includegraphics[width=.7\linewidth]{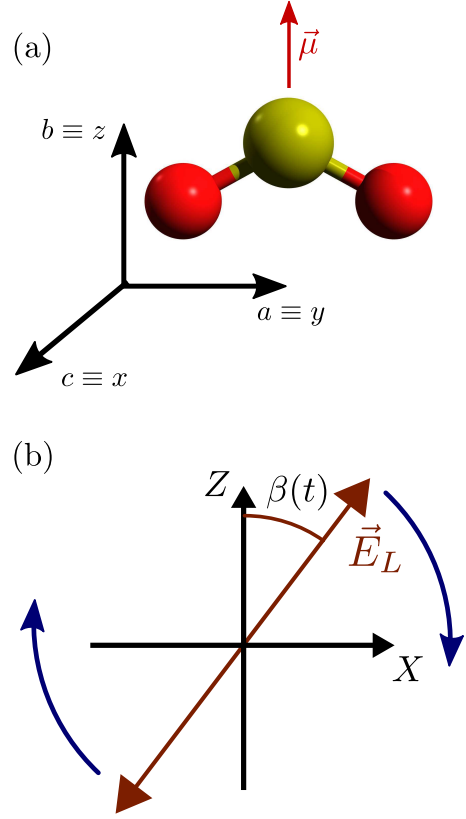}
  \caption{\label{fig:fig1} Sketches of (a) the SO$_2$ molecule referred to the molecular fixed frame; and (b) the polarization axis of the electric field of the laser pulse, $\vec{E}_L$, with respect to the axes of the LFF.}
\end{figure}

We analyze the impact of a static dc field and an intense non-resonant centrifuge laser pulse on the rotational dynamics of SO$_2$. We work in the Born-Oppenheimer and the rigid rotor approximations, neglecting transitions among electronic or vibrational levels. In this framework, the wavefunction is written in terms of the Euler angles $\Omega=(\phi,\,\theta,\,\chi)$, which determine the relative orientation of the $xyz$ molecular fixed frame (MFF) with respect to the $XYZ$ laboratory fixed frame (LFF)~\cite{Zare1988}. In the Born-Oppenheimer and the rigid rotor approximation, the Hamiltonian reads as
\begin{equation}
\label{eq:hamiltonian}
\mathbf{H}(t)=\mathbf{H}_{rot}+\mathbf{H}_{S}+\mathbf{H}_{L}(t),
\end{equation}
where $\mathbf{H}_{rot}$ is the rotational kinetic term
\begin{equation}
\label{eq:hrot}
\mathbf{H}_{rot}=\hbar^{-2}\left(C\mathbf{J_x}^2+A\mathbf{J_y}^2+B\mathbf{J_z}^2\right),
\end{equation}
$\mathbf{J}_k$ being the projection of the angular momentum operator along the $k$ axis of the MFF. The rotational constants are $A=\SI{2.028}{\energy}, B=\SI{.3442}{\energy}~\text{and}~C=\SI{.2935}{\energy}$~\cite{Kallush2015}, where the $a$ axis of the MFF is parallel to the line which contains the oxygen atoms, $b$ lies in the plane of the molecule and contains the sulfur atom and $c$ is perpendicular to the molecular plane [see Fig.~\ref{fig:fig1}(a)]. Note that we identify the axes $a,\, b$ and $c$ with $y,\, z$ and $x$, respectively, throughout the paper. 

The molecule interacts with the static electric field by means of its permanent dipole moment
\begin{equation}
\label{eq:hs}
\mathbf{H}_S=-\vec\mu \cdot \vec{E}_S=-\mu E_S\cos\theta_{Zz},
\end{equation}
where $\vec{E}_S=E_S\hat{Z}$ is the electric field, the dipole moment $\mu=1.62$~D points to the sulfur atom~\cite{Kallush2015} [see Fig.~\ref{fig:fig1}(a)] and $\theta_{Pq}$ is the angle formed by the $P$ and $q$ axes of the LFF and the MFF, respectively. Without loss of generality, we only consider electric fields parallel to the $Z$ axis of the LFF, because the inclination angle can be included in the angle formed by the polarization of the laser field and the $Z$ axis of the LFF. 

The coupling with the non-resonant laser field, $\mathbf{H}_L(t)$ is obtained after averaging over rapid oscillations of its electric field~\cite{stapelfeldt:rev_mod_phys_75_543},
\begin{eqnarray}
\label{eq:hl}
\nonumber
&&\mathbf{H}_{L}(t)=-\frac{1}{4}\vec{E}_L^\dagger(t) \tensornote{\alpha}\vec{E}_L(t)=-\frac{E_L(t)^2}{4}\left\{\alpha_{xx}+\right.\\
\nonumber
&& \cos^2\beta(t)\left[\left(\alpha_{zz}-\alpha_{xx}\right)\cos^2\theta_{Zz}+\left(\alpha_{yy}-\alpha_{xx}\right)\cos^2\theta_{Zy}\right]+\\
\nonumber
&& \sin^2\beta(t)\left[\left(\alpha_{zz}-\alpha_{xx}\right)\cos^2\theta_{Xz}+\left(\alpha_{yy}-\alpha_{xx}\right)\cos^2\theta_{Xy}\right]+\\
\nonumber
&& \sin 2\beta(t) \left[\left(\alpha_{zz}-\alpha_{xx}\right)\cos\theta_{Xz}\cos\theta_{Zz}+\right.\\
\label{eq:coupling_laser_expanded_def}
&&\left.\left.\left(\alpha_{yy}-\alpha_{xx}\right)\cos\theta_{Xy}\cos\theta_{Zy}\right]\right\}
\end{eqnarray}
where   the polarizability tensor, $\tensornote{\alpha}$, is diagonal in the MFF with components $\alpha_{xx}=\SI{2.756}{\pol},\,\alpha_{yy}=\SI{4.638}{\pol}~\text{and}~\alpha_{zz}=\SI{3.082}{\pol}$~\cite{Xenides2000} and $\vec{E}_L(t)$ is the envelope of the pulse. In Appendix~\ref{sec:derivation_of_the_laser_term} we describe in detail the expansion of $\mathbf{H}_{L}(t)$ in terms of the relative orientation between the MFF and the LFF.

A field-dressed state is labeled as the field-free state $J_{K_a,K_c}M$ adiabatically connected with it~\cite{king_jcp11}, where $J$ is the total angular momentum, $M$ the magnetic quantum number and $K_a$ and $K_c$ the projection of the total angular momentum along the $a$ and $c$ axis in the prolate and oblate limiting cases, respectively.
We consider the following process: Initially a SO$_2$ molecule interacts with a static dc field parallel to the $Z$ axis of the LFF. The field is switched on adiabatically until a maximum $E_S$ at $t=0$, and is kept constant. We assume that this process is adiabatic, hence, the wavefunction at $t=0$ is an eigenstate of the field-dressed Hamiltonian $\mathbf{H}_{rot}-\mathbf{H}_S$. In a second step, at $t=0$ an optical centrifuge pulse contained in the $XZ$ plane of the LFF is switched on. Its analytical form is $\vec{E}_L(t)=E_{L,\text{max}}g(t)\left[\hat{X}\sin \beta(t)+\hat{Z}\cos\beta(t)\right]$, where $\beta(t)=\frac{\gamma}{2}t^2+\delta$, $\gamma$ is the angular acceleration of the polarization axis, $\delta$ the angular initial phase and the envelope $g(t)$ reads
\begin{equation}
\label{eq:el_envelope}
g(t)=\left\{
\begin{array}{ll}
\sin^2\left[\cfrac{\pi t}{2 t_\text{on}}\right],&  0\le t \le t_\text{on},\\
&\\
1,& t_\text{on}\le t \le t_\text{p}-t_\text{off},\\
&\\
\sin^2\left[\cfrac{\pi (t-t_\text{p})}{2 t_\text{off}}\right],&  t_\text{p}-t_\text{off}\le t \le t_\text{p},\\
0,& t_\text{p}<t.
\end{array}
\right.
\end{equation}
 In the present work, we consider a pulse with an intensity $\Ialign=\SI{5e12}{\intensity}$, turning on/off times $t_\text{on}=t_\text{off}=\SI{2}{\pico\second}$ and a duration of $t_\text{p}=\SI{32}{\pico\second}$, for several values of $\gamma$. Finally, we also analyze the rotational dynamics in the static electric field once the pulse is switched off.

To investigate the rotational dynamics, we calculate the time dependent  wavefunction by solving the TDSE using the short iterative Lanczos method~\cite{mctdh,Omiste2016_asymmetric_molecules}. For the angular degrees of freedom we use a basis set expansion in terms of the symmetrized field-free symmetric top eigenstates $\ket{JKMs}$~\cite{Omiste2016_asymmetric_molecules}, where $J$ is the total angular momentum quantum number, $K$ and $M$ are its projections along the $z$ axis of the MFF and the $Z$ axis of the LFF, respectively, and $s=0,\, 1$ is the parity under reflections on the polarization plane $XZ$ of the LFF, $\sigma_{XZ}$. The total Hamiltonian $\mathbf{H}(t)$ commutes with  $\sigma_{XZ}$ and two-fold rotation around the $z$-axis of the MFF ($C_2^z$), therefore $s$ and the parity of $K$ are preserved and define the four irreducible representations of the system~\cite{Omiste2011a,Omiste2016_asymmetric_molecules}. See Appendix~\ref{sec:coupling_wigner} for further details on $\ket{JKMs}$ and its relation with the Wigner elements, $D_{M,K}^J(\Omega)$~\cite{Zare1988}. 

\section{Results}
\label{sec:results}
In this section we investigate in detail the rotational dynamics induced in a SO$_2$ molecule by a centrifuge laser pulse and a static dc field. Throughout this work, we only consider even wavefunctions under the symmetry operations $C_{2}^z$ and $\sigma_{XZ}$. The calculations are converged for basis set functions with $0\le J\le J_\text{max}=90$.

\subsection{Centrifugal dynamics}
\label{sec:centrifugal}
\begin{figure}[h]
  \includegraphics[width=.95\linewidth]{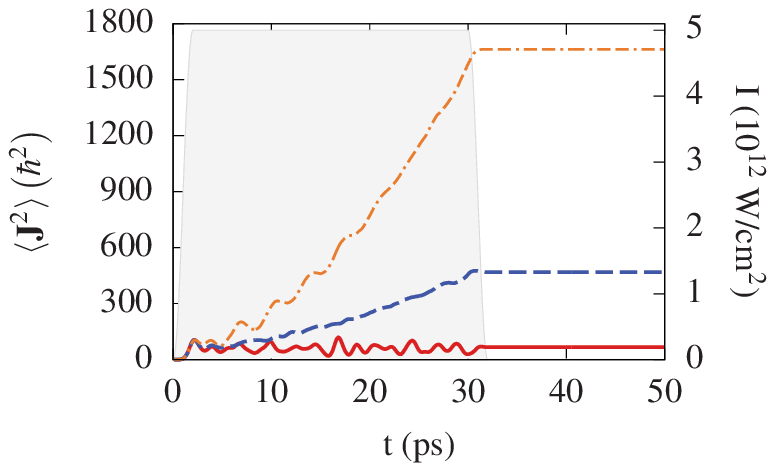}
  \caption{\label{fig:fig2} For the initial state $0_{00}0$, expectation value $\expected{\mathbf{J}^2}$ for an electric field with strength $E_S=\SI{300}{\fieldstrength}$ and $\gamma=\SI{0}{}~\text{(solid red)},~\SI{5}{}$~\text{(dashed blue)}~and~$\SI{10}{\degree/\pico\second^2}$~(dash-dotted orange). The envelope of the centrifuge is also shown (gray).}
\end{figure}

Here we analyze the centrifugal dynamics of the molecule induced by the optical centrifuge with $\delta=0$. The slow rotation of the polarization axis during the turning-on of the laser allows the most polarizable axis (MPA) of the SO$_2$ molecule to align along it. Next, the accelerated rotation of the laser polarization axis is followed by the MPA. After the pulse is over, the molecule continues rotating, ideally, at the final angular frequency of the pulse. This rotation leads to the mixing of the angular momentum of the system. To illustrate this effect, we show in Fig.~\ref{fig:fig2} the expectation value $\expected{\mathbf{J}^2}$ for the rotational groundstate, $0_{00}0$, and $\Ialign=\SI{5e12}{\intensity}$, $E_\text{S}=\SI{300}{\fieldstrength}$ and $\gamma=0,\,5~\text{and}~\SI{10}{\degree/\pico\second^2}$. For $\gamma=0$, $\expected{\mathbf{J}^2}$ increases until $66.93\hbar^2$ and oscillates around this value until the pulse is off, remaining almost constant.

 Just after the turning-on of the laser, the optical centrifuge is still slow, therefore, $\expected{\mathbf{J}^2}$ almost coincides for $\gamma=\SI{0}{},\,\SI{5}{}~\text{and}~\SI{10}{\degree/\pico\second^2}$, up to the first peak of approximately $105\hbar^2$ around 2.17 ps. However, as the polarization axis accelerates, the angular momentum increases linearly in time, tending to be proportional to the angular velocity. This effect can be understood classically for a linear rotor, where the classical angular momentum is $J_\text{cl}=I\omega$, being $I$ the inertia constant and $\omega$ the angular velocity which, in this setup, increases linearly in time. Under this assumption, we may approximate at the end of the pulse $\frac{J^2_\text{cl}{(\gamma=\SI{10}{\degree/\pico\second^2})}}{J^2_\text{cl}{(\gamma=\SI{5}{\degree/\pico\second^2})}}\approx\frac{10^2}{5^2}=4$, which is close to the result of the time propagation $\frac{\expected{\mathbf{J^2}}_{\gamma=\SI{10}{\degree/\pico\second^2}}}{\expected{\mathbf{J^2}}_{\gamma=\SI{5}{\degree/\pico\second^2}}}=\frac{1662.41}{468.92}\approx 3.54$.

\begin{figure}[h]
  \includegraphics[width=.95\linewidth]{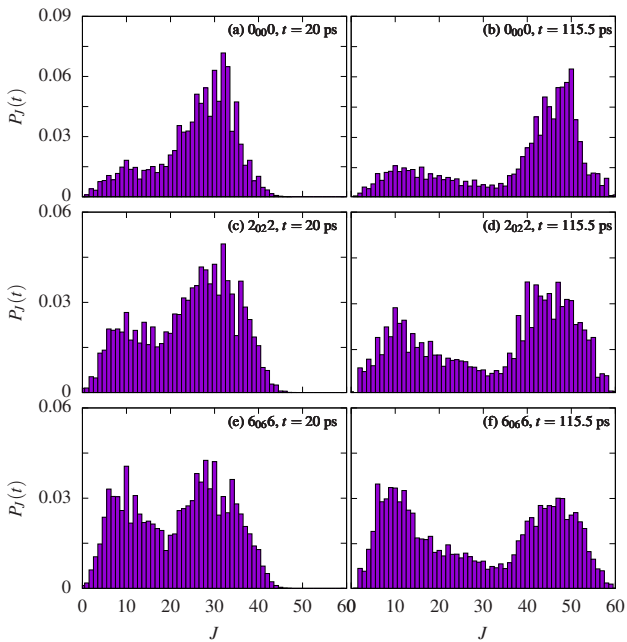}
  \caption{\label{fig:fig3} For the initial states $0_{00}0$, $2_{02}2$ and $6_{06}6$, the distribution of the coefficients with angular momentum $J$, $P_J(t)$, for an electric static field with strength $E_S=\SI{300}{\fieldstrength}$ and an acceleration of the centrifuge $\gamma=\SI{10}{\degree/\pico\second^2}$ at $t=20$~ps (a), (c) and (e) and $t=115.5$~ps (b), (d) and (f).}
\end{figure}

To obtain a better physical insight of the rotational dynamics of the wavefunction $\Psi(t)$ we analyze the population of all the components with the same rotational quantum number $J$ but different values of $M$ and $K$, which is defined as $P_J(t)=\sum_{KMs}|\escalar{JKMs}{\Psi(t)}|^2$. 
In Fig.~\ref{fig:fig3} we show the distribution of $P_J(t)$ during and after the pulse for the initial states ${J}_{{K_a},{K_c}}{M}=0_{00}0$, $2_{02}2$ and $6_{06}6$. First note that the weights of even and odd $J$ contributions to $P_J(t)$ are comparable due to the mixing induced by the dc electric field during the turning-on of the laser, as has already been proven for excitations by a non-resonant linearly polarized laser~\cite{Nielsen2012,omiste:pra2012,Omiste2016_asymmetric_molecules}. However, there is still a predominance of the even contributions over the odd. For each initial state, we observe that $P_J(t=20\text{~ps})$ is formed by two distributions with a large overlap, centered approximately around $J=8\text{~and~}J=28$, being more remarkable for $2_{02}2$ and $6_{06}6$. After the pulse, the distribution at low $J$ has slightly changed not only the shape, but also the population. This part of the distribution is constituted by the components which are unable to follow the rotation of the laser polarization axis. However, the other part is pushed to larger $J$ values during the acceleration, increasing the net rotation of the molecule. For the initial state $6_{06}6$ the population of the components which does not contribute to the rotation is larger, since the higher angular momentum of the initial state implies that more counter rotating components play a role during the dynamics. For the groundstate as initial state $\expected{\vec{\mathbf{J}}^2}\approx 0$ before the turning-on of the laser, which ensures that the initial population is not corotating or counterrotating.

\begin{figure}[h]
  \includegraphics[width=.95\linewidth]{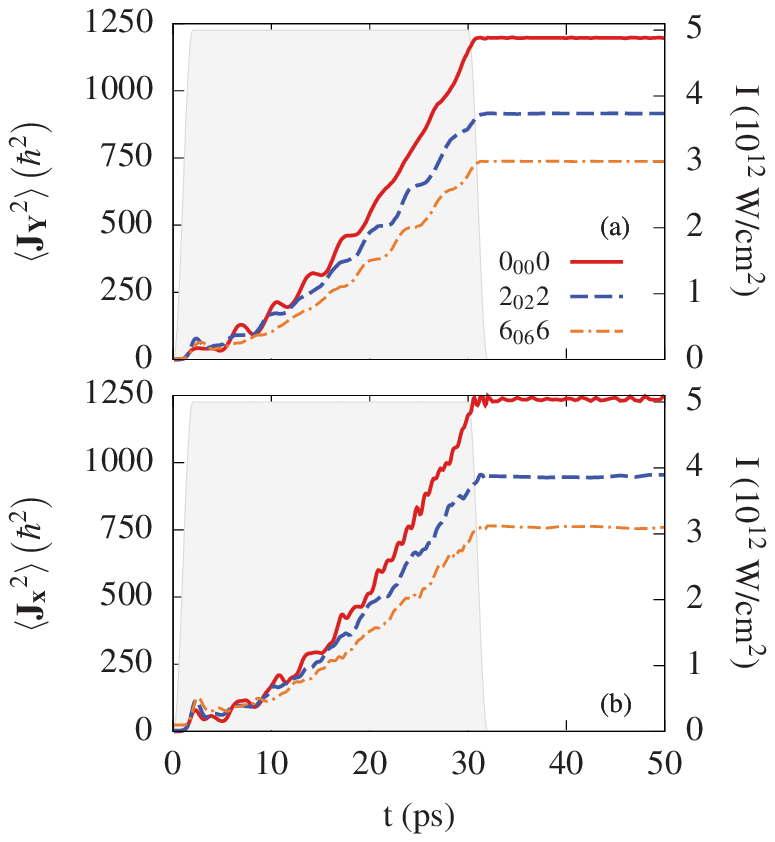}
  \caption{\label{fig:fig4} For the initial states $0_{00}0$ (red solid), $2_{02}2$ (dashed blue) and $6_{06}6$ (orange dash-dotted) expectation values (a) $\expected{\mathbf{J_Y}^2}$ and (b) $\expected{\mathbf{J_x}^2}$ as a function of $t$ for an angular acceleration $\gamma=\SI{10}{\degree/\pico\second^2}$ and an electric field with strength $E_S=\SI{300}{\fieldstrength}$. The envelope of the centrifuge is also shown (gray).}
\end{figure}

We now analyze the rotational dynamics of the molecule in terms of the projection of the angular momentum $\vec{\mathbf{J}}$ along the axes of the MFF and the LFF. The optical centrifuge induces a rotation around the $Y$ axis of the LFF, which is measured by $\expected{\mathbf{J_Y}^2}$, shown in Fig.~\ref{fig:fig4}(a). $\expected{\mathbf{J_Y}^2}$ follows the same behavior as $\expected{\mathbf{J}^2}$, increasing during the pulse, until a maximum value which depends on the initial state. During the pulse, we observe oscillations which also appear in the acceleration of linear rotors~\cite{Spanner2001a}. After the pulse, $\expected{\mathbf{J_Y}^2}$ remains almost constant, being the weak dc field responsible for the tiny oscillations. In Fig.~\ref{fig:fig4}(b) we show the square of the projection of $\vec{\mathbf{J}}$ along the $x$-axis of the MFF, $\expected{\mathbf{J_x}^2}$, for the three initial states. We observe that $\expected{\mathbf{J_Y}^2}\approx\expected{\mathbf{J_x}^2}$ during and after the optical centrifuge, despite the oscillation of $\expected{\mathbf{J_x}^2}$ caused  by the asymmetric inertia tensor and the disagreements at low angular momentum. Moreover, these two projections constitute the major contribution to the total angular momentum. For instance, at the end of the pulse, $\expected{\mathbf{J}^2}\approx 1662.41\hbar^2$ for the groundstate at $\gamma=\SI{10}{\degree/\pico\second^2}$, whereas $\expected{\mathbf{J_x}^2}\approx 1235.51\hbar^2$. Therefore, the molecule tends to restrict the motion of the MPA ($y$-axis of the MFF) and the remaining axis with the largest rotational constant ($z$ axis) in the plane of the optical centrifuge ($XZ$ plane), as has been shown experimentally~\cite{Korobenko2015a}.

Taking all this into account, we are allowed to approximate the superrotor after the pulse as a linear rotor, whose rotational constant is $C$~\cite{Korobenko2015a}. Then, we can extract the \emph{effective} angular velocity $\omega_{ef}=\sqrt{\expected{\omega^2}}$ using $\expected{\mathbf{J_x}^2}=I_C^2\expected{\omega^2}$, where $I_C$ is the inertia constant around the $x$ axis and $C=\hbar^2/(2I_C)$. We obtain $\omega_{ef}= 221.86,\,195.14,~\text{and}~\SI{174.23}{\degree/\pico\second}$ for $0_{00}0$, $2_{02}2$ and $6_{06}6$, respectively. Note that $C$ is the smallest rotational constant, hence, these values are lower bounds of the real $\omega$, which would correspond to $\SI{320}{\degree/\pico\second}$ and $\expected{\mathbf{J_x}^2}\sim\expected{\mathbf{J}^2}=50.51\hbar^2$. Let us remark that the peak of $P_J(t)$ for high $J$'s after the pulse is located around $J\approx 50$ in Fig.~\ref{fig:fig3}~(b), (d) and (f), in agreement with the approximation of SO$_2$ as a linear rotor.

\subsection{Alignment to the polarization plane}
\label{sec:alignment_polarization_plane}

\begin{figure}[h]
  \includegraphics[width=.95\linewidth]{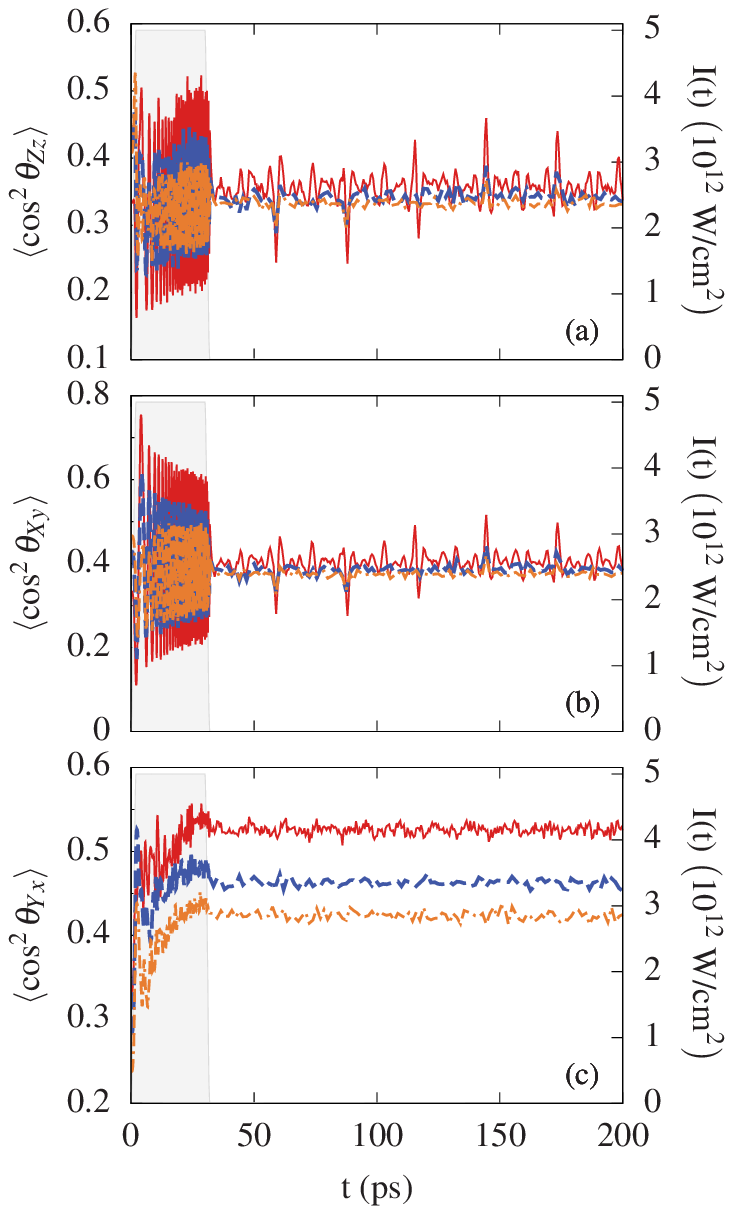}
  \caption{\label{fig:fig5} For the initial states $0_{00}0$ (red solid), $2_{02}2$ (blue dashed) and $6_{06}6$ (orange dash-dotted), the expectation values (a) $\expected{\cos^2\theta_{Zz}}$, (b) $\expected{\cos^2\theta_{Xy}}$ and (c) $\expected{\cos^2\theta_{Yx}}$ for an electric field with strength $E_S=\SI{300}{\fieldstrength}$ and a angular acceleration $\gamma=\SI{10}{\degree/\pico\second^2}$. The envelope of the centrifuge is also shown (gray).}
\end{figure}

Next, we analyze the alignment induced by the optical centrifuge in the LFF. In the previous
section, the comparison between $\expected{\mathbf{J_x}^2}~\text{and}~\expected{\mathbf{J_Y}^2}$
showed that after the optical centrifuge most of the rotation of the SO$_2$ molecule is restricted
around the optical centrifuge propagation axis and the $x$ axis of the MFF. Therefore, the $x$ axis of the MFF tends to align along the $Y$ axis of the LFF,
that is to say, the plane defined by the molecule leans towards the $XZ$ plane of the LFF. This
effect has been observed experimentally in both linear~\cite{Milner2016a} and asymmetric top
molecules~\cite{Korobenko2015a}. To illustrate the alignment of the molecule we show the alignment factors
$\expected{\cos^2\theta_{Zz}}$, $\expected{\cos^2\theta_{Xy}}$ and $\expected{\cos^2\theta_{Yx}}$
for the initial states $0_{00}0$, $2_{02}2$ and $6_{06}6$, and the pulse parameters $\Ialign=\SI{5e12}{\intensity}$, $\gamma=\SI{10}{\degree/\pico\second^2}$ and
$E_S=\SI{300}{\fieldstrength}$ in Fig.~\ref{fig:fig5}. During the turning on of the laser, the polarization axis is
almost parallel to the $Z$ axis of the LFF. Hence, the MPA aligns along it, and thus $\expected{\cos^2\theta_{Xy}}$ decreases to
approximately $0.110$ for $0_{00}0$. Similarly, $\expected{\cos^2\theta_{Zz}}$
reaches a minimum of approximately $0.168$ at the same times. Due to the optical centrifuge, the
projections of these molecular axes onto the lab axes oscillate following the polarization axis rotation. 
For the three initial states, $\expected{\cos^2\theta_{Zz}}$ and
$\expected{\cos^2\theta_{Xy}}$ follow the same pattern during the pulse, but the amplitude of the
oscillations diminishes as the excitation of the state increases. After the pulse, each alignment factor oscillates around approximately the same value for all the states considered, being a bit larger for $\expected{\cos^2\theta_{Xy}}$, which involves the MPA. The revivals for both alignment factors during the post pulse propagation are located at the same times. They are more marked for the rotational groundstate as initial state, where the maximum peaks for $\expected{\cos^2\theta_{Zz}}$ and $\expected{\cos^2\theta_{Xy}}$ are 0.459 and 0.516 and are located at $t\approx 144.5$~ps. The alignment factors are mainly driven by C-type revivals~\cite{Felker1992,Tenney2016}, being the revival time $T_\text{rev}=57$~ps, in accordance to the experimental measurements~\cite{Korobenko2015a} and the rotational dynamics described in Sec.~\ref{sec:centrifugal}. 
In Fig.~\ref{fig:fig5}(c) we illustrate the alignment in the $XZ$ plane of the LFF by $\expected{\cos^2\theta_{Yx}}$. Let us remark that $0\le \expected{\cos^2\theta_{Yx}}\le 1$, where $0$ means that the molecular plane is perpendicular to the $XZ$ plane of the LFF and, on the contrary, they are coplanar for $\expected{\cos^2\theta_{Yx}}=1$. The alignment dynamics is similar for the three states, but, as expected, it is more efficient for the $0_{00}0$. During the turning on, the alignment increases abruptly and the rotation of the polarization axis induces a smooth increasing until the pulse is over. The turning-off of the laser field weakly affects the maximum value reached during the centrifuge, corresponding to approximately $0.528,\,0.463~\text{and }0.424$ for $0_{00}0$, $2_{02}2$ and $6_{06}6$, respectively. These values remain during the post pulse propagation, \ie, SO$_2$ remains attached to the $XZ$ plane of the LFF for long times due to angular momentum conservation and the stability of rotations around the $x~(c)$ axis.

\begin{figure}[h]
  \includegraphics[width=.95\linewidth]{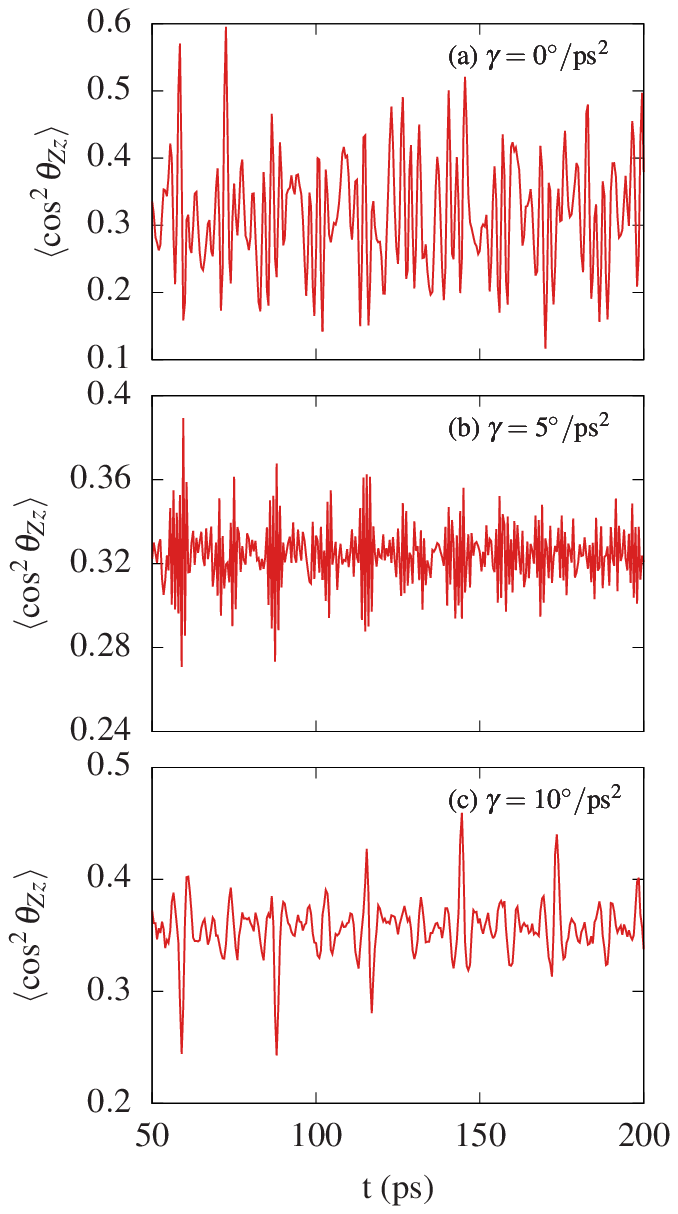}
  \caption{\label{fig:fig6} For the initial state $0_{00}0$, the expectation value $\expected{\cos^2\theta_{Zz}}$ for an electric field with strength $E_S=\SI{300}{\fieldstrength}$, a centrifuge with peak intensity $\Ialign=\SI{5e12}{\intensity}$ and (a) $\gamma=\SI{0}{}$, (b) $\SI{5}{}$ and (c) $\SI{10}{\degree/\pico\second^2}$.}
\end{figure}

Finally, in Fig.~\ref{fig:fig6}, we show the post pulse dynamics of the alignment $\expected{\cos^2\theta_{Zz}}$ of the groundstate for  the angular accelerations $\gamma=0,\,5~\text{and }\SI{10}{\degree/\pico\second^2}$. For $\gamma=\SI{0}{\degree/\pico\second^2}$,~\ie, a linearly polarized laser pulse, the alignment presents an irregular behavior without any pattern, ranging from $0.15$ to $0.43$. For $\gamma=\SI{5}{\degree/\pico\second^2}$ we observe some revivals separated by approximately $28.4$~ps.  There are also weaker revival-like structures between two consecutive main revivals, due to the contribution of other rotation modes associated to SO$_2$. Note that the asymmetry of the SO$_2$ molecule implies that these revivals are not well defined, hence, the interference of different modes causes the damping and vanishing of these structures at long times. For faster rotations, as in the case of $\gamma=\SI{10}{\degree/\pico\second^2}$, the main revivals are more pronounced, because the rotation around the $x$ axis of the molecule prevails over the other motions.

\subsection{Orienting superrotors}
\label{sec:orienting_superrotors}

In this section we analyze the orientation induced by the impact of the dc field and the optical centrifuge. Let us recall that the dipole moment of the SO$_2$ molecule defines the $z$ axis of the MFF and coincides with the second MPA. On the contrary, the MPA lies along the $y$ axis of the MFF. Taking into account that even for the largest acceleration considered, $\gamma=\SI{10}{\degree/\pico\second^2}$, the polarization axis only rotates $20\degree$ during the turning on, we can restrict our study to two limiting cases for linearly polarized lasers ($\gamma=\SI{0}{\degree/\pico\second^2}$) to understand the dynamics during the switching on of the centrifuge. First, for parallel fields ($\delta=0\degree$), the non-resonant laser pushes the MPA to its polarization axis. However, the dc field acts in the opposite way, forcing the dipole moment $\vec{\mu}$ to orient along the same axis of the LFF. For weak dc fields, the orientation is negligible due to the strong interaction due to the laser. On the other hand, in the perpendicular case ($\delta=90\degree$), both fields collaborate and the orientation along the dc field is compatible with the alignment along the polarization axis of the laser. In addition to these considerations, the field configuration determines the number of real and avoided crossings as well as the population transfer among them~\cite{Omiste2013,Omiste2016_asymmetric_molecules}, which can dramatically affect the orientation, even for weak dc fields~\cite{Omiste2011, Nielsen2012}. 
\begin{figure}[h]
  \includegraphics[width=.95\linewidth]{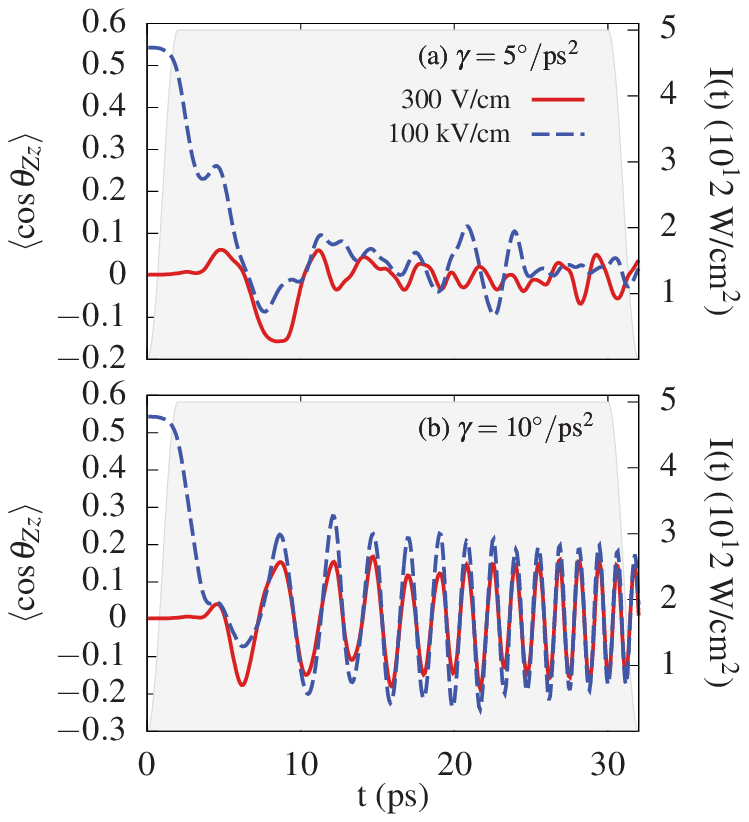}
  \caption{\label{fig:fig7} For the initial state $0_{00}0$, expectation value $\expected{\cos\theta_{Zz}}$ during the interaction with the optical centrifuge for an electric static field with strength $E_S=\SI{300}{\fieldstrength}$ (solid red) and $\SI{100}{\kfieldstrength}$ (dashed blue),  an angular acceleration (a) $\gamma=\SI{5}{}$ and (b) $\SI{10}{\degree/\pico\second^2}$ and an initial angle of the centrifuge  $\delta=90\degree$. The envelope of the centrifuge is also shown (gray).}
\end{figure}

We illustrate the orientation of the groundstate in the presence of the optical centrifuge for $\delta=\SI{90}{\degree}$ in Fig.~\ref{fig:fig7}. Before the centrifuge is switched on, the orientation $\expected{\cos\theta_{Zz}}=\SI{.234e-2}{}$ and $0.543$ for $E_S=\SI{300}{\fieldstrength}$ and $\SI{100}{\kfieldstrength}$, respectively. The fast switching on of the laser constructs a coherent wavepacket which enables the orientation and antiorientation during the propagation, as it is observed even for $\gamma=\SI{0}{\degree/\pico\second^2}$. We see in Fig.~\ref{fig:fig7}(b) that the orientation is fully controlled by the laser field for $\gamma=\SI{10}{\degree/\pico\second^2}$, and coincide for both strengths of the dc field. However, for $\gamma=\SI{5}{\degree/\pico\second^2}$ the laser is not able to drive the orientation, as we show in Fig~\ref{fig:fig7}(a). 

\begin{figure}[h]
  \includegraphics[width=.95\linewidth]{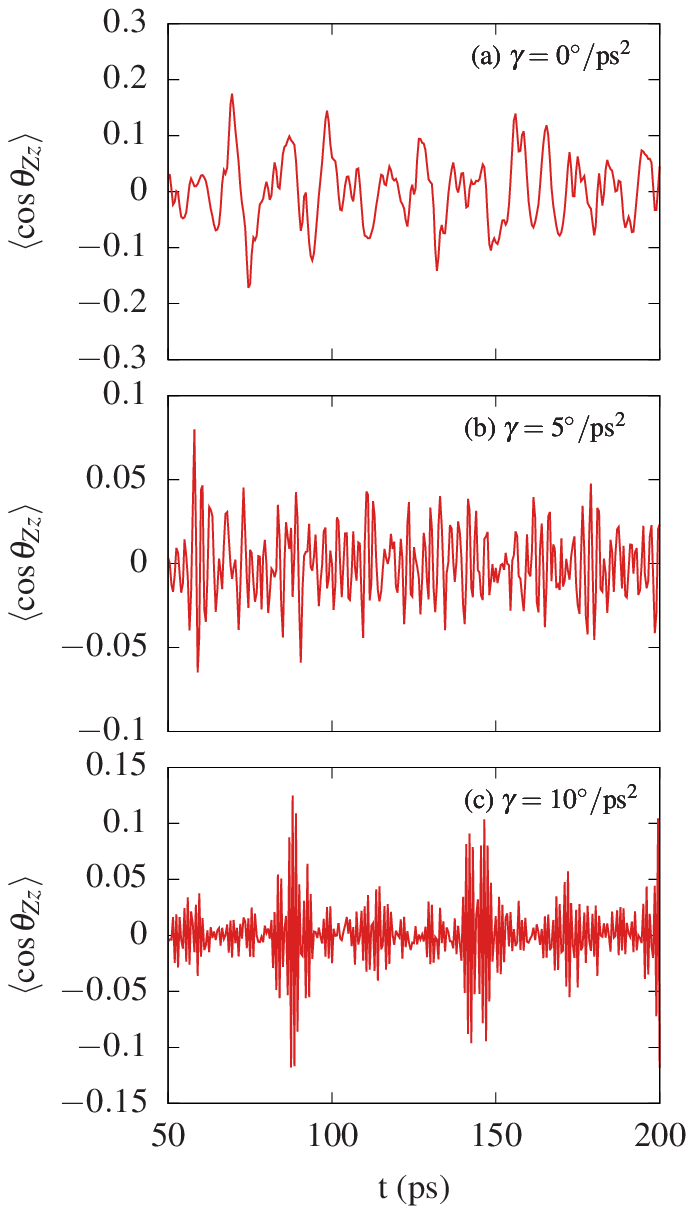}
  \caption{\label{fig:fig8} For the initial state $0_{00}0$, expectation value $\expected{\cos\theta_{Zz}}$ for an electric field with strength $E_S=\SI{300}{\fieldstrength}$, angular accelerations (a) $\gamma=\SI{0}{}$, (b) $\SI{5}{}$ and (c) $\SI{10}{\degree/\pico\second^2}$ and an initial angle of the centrifuge $\delta=\SI{90}{\degree}$.}
\end{figure}

\begin{figure}[h]
  \includegraphics[width=.95\linewidth]{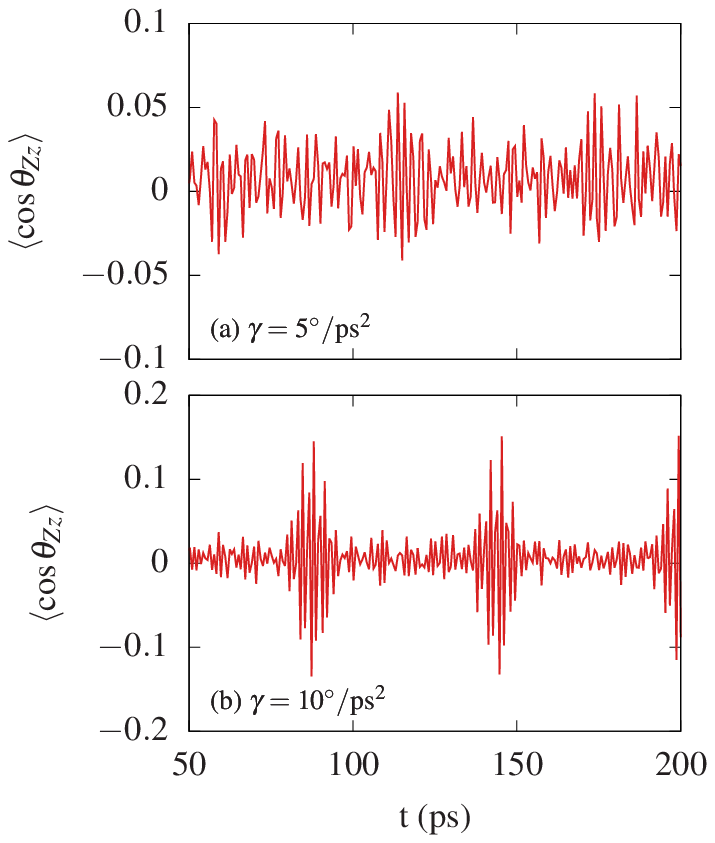}
  \caption{\label{fig:fig9} For the initial state $0_{00}0$, expectation value $\expected{\cos\theta_{Zz}}$ for an electric field with strength $E_S=\SI{100}{\kfieldstrength}$, angular accelerations (a) $\gamma=\SI{5}{}$ and (b) $\SI{10}{\degree/\pico\second^2}$ and an initial angle of the centrifuge $\delta=\SI{90}{\degree}$.}
\end{figure}

\begin{figure}[h]
  \includegraphics[width=.95\linewidth]{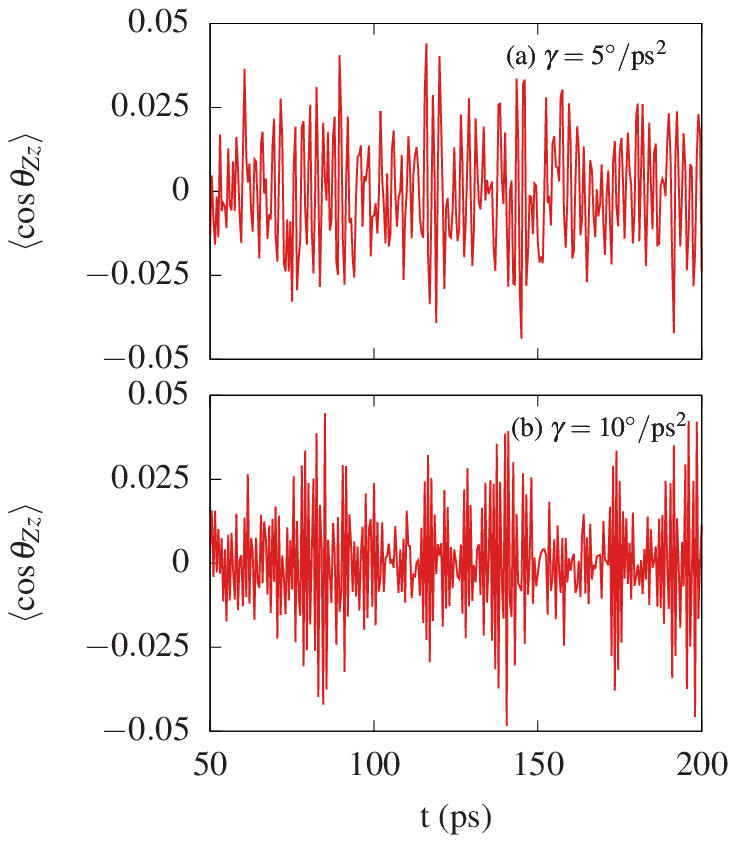}
  \caption{\label{fig:fig10} For the initial state $0_{00}0$, expectation value $\expected{\cos\theta_{Zz}}$ for an electric field with strength $E_S=\SI{300}{\fieldstrength}$, angular accelerations (a) $\gamma=\SI{5}{}$ and (b) $\SI{10}{\degree/\pico\second^2}$ and an initial angle of the centrifuge $\delta=\SI{0}{\degree}$.}
\end{figure}

These considerations have important implications in the post pulse propagation, as we illustrate in Figs.~\ref{fig:fig8}-\ref{fig:fig10}. First, let us evaluate the impact of the remaining dc field during the rotation of the molecule after the pulse. We consider that the kinetic energy of the superrotor is mainly due to the rotation around the $x$ axis of the MFF [see Sec.~\ref{sec:centrifugal}], then $\expected{\mathbf{H}_{rot}}\sim \hbar^{-2}C\expected{\mathbf{J}_x^2}\approx \SI{132.07}{\energy}$ for $\gamma=\SI{5}{\degree/\pico\second^2}$, which is much larger than the coupling with $E_S=\SI{100}{\kfieldstrength}$, $\sim E_S\expected{\mu_z}\approx \expected{\cos\theta_{Zz}}\times \SI{2.72}{\energy}$. Therefore, during the rotation, the superrotor will experience a negligible deceleration (acceleration) impulse when oriented (antioriented) with respect to the electric field. Let us remark that the impact of this kick on the orientation is even much smaller for $\gamma=\SI{10}{\degree/\pico\second^2}$, since $\expected{\mathbf{H}_{rot}}\sim \hbar^{-2}C\expected{\mathbf{J}_x^2}\approx \SI{366.87}{\energy}$.
In Fig.~\ref{fig:fig8} we show the orientation for the most favorable case, \ie, $\delta=90\degree$. For $\gamma=\SI{0}{\degree/\pico\second^2}$ we observe an irregular orientation pattern ranging from approximately $-0.2~\text{to }0.2$, with decreasing amplitude over time. If we increase the acceleration to $\SI{5}{\degree/\pico\second^2}$ the peak orientation decreases to 0.08 and the oscillatory pattern is irregular. On the contrary, the orientation for $\gamma=\SI{10}{\degree/\pico\second^2}$ shows a clear and well-defined revival structure characterized by revivals located around $85,140$ and $198$~ps,  which are separated by approximately $T_\text{rev}=57$~ps, associated to the rotations around the $x$ axis of the MFF. Between these revivals we find other oscillations which are related to other rotational motions.
 
 If we increase the dc field strength to $E_S=\SI{100}{\kfieldstrength}$ the revivals in the orientation become more regular, as we see in Fig.~\ref{fig:fig9}(b). Specifically, the location of the revivals are separated again by approximately $57$~ps, but, in contrast to the previous case, there are no revivals between these structures. The oscillations of the main revivals in the laser field free region  are slightly enhanced with respect to the weak dc scenario. 

We illustrate the collinear fields case with a weak dc field, $E_S=\SI{300}{\fieldstrength}$ in Fig.~\ref{fig:fig10}. As we have discussed above, the dc field and the laser field attempt to orient and align the molecular axes along different directions during the switching on. For $\gamma=\SI{5}{\degree/\pico\second^2}$ we find that the orientation is highly oscillatory without a main frequency or revival structure as in $\delta=\SI{90}{\degree}$ and the amplitude is smaller than 0.05. However, for $\gamma=\SI{10}{\degree/\pico\second^2}$ we do not recover a well-defined revival structure, caused by the mixing during the switching on of the centrifuge. 

As we have discussed, the initial configuration of the fields, \ie, $\delta$ and the strength of the field, has a strong impact in the orientation for the parameters analyzed in this work. However, for higher values of $\gamma$, the angle covered during the switching on may exceed $\pi$, increasing the population transfer among states with different orientation. This may lead to a vanishing average value of the orientation. 

\section{Conclusions}
\label{sec:conclusions}

We have theoretically studied the rotational dynamics of an asymmetric rotor induced by an optical centrifuge and a constant static electric field by solving the TDSE in the rigid-rotor approximation. Specifically, we describe the case of sulfur dioxide, which has been experimentally addressed~\cite{Korobenko2015a}. The accelerated rotation of the polarization axis of the pulse provokes an effective rotation of the molecule paced with the centrifuge, which leads to a strong excitation of states with high angular momentum. We observe that the population of $J$ as the function of time is formed by two well defined parts: the efficiently accelerated and the non accelerated. The latter one is characterized by $J\sim 10$ for $\gamma=\SI{10}{\degree/\pico\second^2}$ and remains almost unaltered during the acceleration, whereas the accelerated part moves to higher $J$'s as the centrifuge accelerates. The accelerated population is lower for higher excited initial states.

For the first time, we have shown numerically that a planar asymmetric molecule tends to the plane defined by the polarization axis of the centrifuge as the superrotor accelerates. We confirm that the superrotor remains attached to this plane for long times after the pulse~\cite{Korobenko2015a}, being slightly affected by the revivals. On the other hand, the delay between revivals in the alignment $\expected{\cos^2\theta_{Zz}}$ and $\expected{\cos^2\theta_{Xy}}$ during the post pulse propagation coincide with the experimental measurements~\cite{Korobenko2015a}.

The restriction of the rotation to the $XZ$ plane implies that the squared of the projection of the angular momentum along the propagation axis of the laser, $\expected{\mathbf{J}_Y^2}$, coincides with the projection along the least polarizable axis (LPA) of the molecule $\expected{\mathbf{J}_x^2}$, being the major contribution to the total angular momentum. Therefore, we can extract the \emph{effective angular velocity} $\omega_{ef}$ using the inertia constant of the rotation around the LPA. Finally, we have analyzed the orientation of the superrotor caused by the dc field. We have shown that for the optical centrifuge accelerations under study, the orientation is very sensitive to the initial angle formed by the polarization axis and the dc field. In the case of SO$_2$, the initial perpendicular configuration is the most favorable for the orientation, since the MPA and the dipole moment are perpendicular. In this scenario, we clearly observe a revival structure, which experimentally may allow to locate the high orientation/antiorientation periods during the time evolution. Moreover, for a strong dc field we observe more well defined revivals than in the weak field case, due mainly to the suppression of many rotational modes which do not correspond to the rotation around the $c$ axis. Let us note that faster rotating optical centrifuges may completely frustrate the orientation during the acceleration.

Summing up, we have shown that the optical centrifuge combined with a static electric field contained in the polarization plane allows to control both the alignment and the orientation of a molecular ensemble of asymmetric molecules. This motivates the exploration of other field configurations such as the combination of an optical centrifuge and a perpendicular electric field, which might produce a large orientation perpendicular to the polarization plane.

\appendix

\section{Derivation of the laser term}
\label{sec:derivation_of_the_laser_term}
The coupling with the laser field in the rotating wave approximation is given by~\cite{stapelfeldt:rev_mod_phys_75_543}
\begin{equation}
  \label{eq:coupling_laser}
  H_L(t)=-\frac{1}{4}\vec{E}^\dagger_L(t) \tensornote{\alpha}\vec{E}_L(t),
\end{equation}
where $\vec{E}_L(t)$ is the envelope of the electric field of the laser field and $\tensornote{\alpha}$ is the polarizability tensor. In the case of SO${}_2$, $\tensornote{\alpha}$ is diagonal in the MFF, with the elements $\alpha_{xx},~\alpha_{yy}~\text{and}~\alpha_{zz}$. The electric field in the MFF reads as
\begin{equation}
  \label{eq:el_mff}
  \vec{E}_L(t)|_\text{\tiny{MFF}}=R(\phi,\theta,\chi)
E_L(t)\left(
  \begin{array}{c}
   \sin\beta(t)\\
    0 \\
    \cos\beta(t)
  \end{array}
\right),
\end{equation}
being $\beta(t)$ the angled formed by the polarization axis of the laser and the $Z$-axis of the LFF. $R(\phi,\theta,\chi)$ is the rotation matrix which links the LFF and the MFF 
\begin{equation}
  \label{eq:rotation_lff_mff}
  R(\phi,\theta,\phi)=\left(
    \begin{array}{ccc}
      \cos\theta_{Xx} & \cos\theta_{Yx} & \cos\theta_{Zx}\\
      \cos\theta_{Xy} & \cos\theta_{Yy} & \cos\theta_{Zy}\\
      \cos\theta_{Xz} & \cos\theta_{Yz} & \cos\theta_{Zz}
    \end{array}
\right),
\end{equation}
where $\theta_{Pq}$ is the angle formed by the $P$ axis of the LFF and the $q$ axis of the MFF.
The analytical expressions of the directional cosines are given in Appendix~\ref{sec:coupling_wigner}. The coupling in Eq.~\eqref{eq:coupling_laser} may be written
\begin{widetext}
  \begin{eqnarray}
    \nonumber
    H_L(t)&=&-\frac{E_L(t)^2}{4}\left[\sin^2\beta(t)\left(\alpha_{xx}\cos^2\theta_{Xx}+\alpha_{yy}\cos^2\theta_{Xy}+\alpha_{zz}\cos^2\theta_{Xz}\right)+\right.\\
\nonumber
&&\cos^2\beta(t)\left(\alpha_{xx}\cos^2\theta_{Zx}+\alpha_{yy}\cos^2\theta_{Zy}+\alpha_{zz}\cos^2\theta_{Zz}\right)+\\
\label{eq:coupling_laser_expanded}
&&\left.2\sin\beta(t)\cos\beta(t)\left(\alpha_{xx}\cos\theta_{Zx}\cos\theta_{Xx}+\alpha_{yy}\cos\theta_{Zy}\cos\theta_{Xy}+\alpha_{zz}\cos\theta_{Zz}\cos\theta_{Xz}\right)\right].
  \end{eqnarray}
\end{widetext}
Using that $\cos\theta_{Xx}\cos\theta_{Zx}+\cos\theta_{Xy}\cos\theta_{Zy}+\cos\theta_{Xz}\cos\theta_{Zz}=0$ and $\cos^2\theta_{Zx}+\cos^2\theta_{Zy}+\cos^2\theta_{Zz}=1$ in Eq.~\eqref{eq:coupling_laser_expanded} we obtain
  \begin{eqnarray}
    \nonumber
    &&\mathbf{H}_L(t)=-\frac{E_L(t)^2}{4}\left\{\alpha_{xx}+\right.\\
\nonumber
&& \cos^2\beta(t)\left[\left(\alpha_{zz}-\alpha_{xx}\right)\cos^2\theta_{Zz}+\left(\alpha_{yy}-\alpha_{xx}\right)\cos^2\theta_{Zy}\right]+\\
\nonumber
&& \sin^2\beta(t)\left[\left(\alpha_{zz}-\alpha_{xx}\right)\cos^2\theta_{Xz}+\left(\alpha_{yy}-\alpha_{xx}\right)\cos^2\theta_{Xy}\right]+\\
\nonumber
&& \sin 2\beta(t) \left[\left(\alpha_{zz}-\alpha_{xx}\right)\cos\theta_{Xz}\cos\theta_{Zz}+\right.\\
\label{eq:coupling_laser_expanded_def_appendix}
&&\left.\left.\left(\alpha_{yy}-\alpha_{xx}\right)\cos\theta_{Xy}\cos\theta_{Zy}\right]\right\}
  \end{eqnarray}

\section{Properties of the Wigner D-matrix elements}
\label{sec:coupling_wigner}

We briefly summarize the properties of the Wigner D-matrix elements used throughout this work. First, the basis set functions of the representations of the total Hamiltonian are given by
\begin{eqnarray*}
&&\ket{JKMs}=\\
&&\left\{
\begin{array}{l}
 \ket{J00},~\text{for}~M,K=0 \\ 
        \\
        \cfrac{1}{\sqrt{2}}\left( \ket{JKM}+(-1)^{M-K+s}\ket{J-K-M}\right),~\text{otherwise},
        \end{array}
        \right.
        \end{eqnarray*} 
where the eigenstates of the symmetric top, $\ket{JKM}$, are written in terms of the Wigner D-matrix elements
\begin{equation}
\label{eq:symmetric_top_basis}
\escalar{\Omega}{JKM}=\sqrt{\cfrac{2J+1}{8\pi^2}}(-1)^{M-K}D_{-M,-K}^J(\Omega).
\end{equation}
The trigonometric functions in $\mathbf{H}_{S}$ and $\mathbf{H}_L(t)$ in expressions~\eqref{eq:hrot} and~\eqref{eq:coupling_laser_expanded_def}, respectively, can also be expressed as linear combinations of $D_{M,K}^J(\Omega)$. The terms involved in $\mathbf{H}_S$
\begin{eqnarray}
  \label{eq:czz}
  &&\cos\theta_{Zz}= \cos\theta=D_{0,0}^1(\Omega)\\
  \label{eq:czx}
  &&\cos\theta_{Xz}= \sin\theta\cos\phi=\frac{1}{\sqrt{2}}\left(D_{-1,0}^1(\Omega)-D_{1,0}^1(\Omega)\right)
\end{eqnarray}
and in $\mathbf{H}_L(t)$
\begin{widetext}
  \begin{eqnarray}
    \label{eq:c2zz}
    \cos^2\theta_{Zz}&=&\cos^2\theta=\frac{1}{3}\left(2D_{0,0}^2(\Omega)+1\right)\\
    \label{eq:c2zy}
    \cos^2\theta_{Zy}&=&\sin^2\theta\sin^2\chi=\frac{1}{3}\left(1-D_{0,0}^2(\Omega)\right)-\sqrt{\frac{1}{6}}\left(D^2_{0,2}(\Omega)+D^2_{0,-2}(\Omega)\right)\\
    \label{eq:c2xz}
    \cos^2\theta_{Xz}&=&\cos^2\phi\sin^2\theta=\frac{1}{3}\left(1-D_{0,0}^2(\Omega)\right)+\sqrt{\frac{1}{6}}\left(D^2_{2,0}(\Omega)+D^2_{2,0}(\Omega)\right)\\
    \nonumber
    \cos^2\theta_{Xy}&=&(-\cos\phi\cos\theta\sin\chi-\sin\phi\cos\chi)^2=-\frac{1}{4}\left(D_{2,2}^2(\Omega)+D_{2,-2}^2(\Omega)+ D_{-2,2}^2(\Omega)+D_{-2,-2}^2(\Omega)\right)+\\
    \label{eq:c2xy}
    &&\sqrt{\frac{1}{24}}\left(D_{0,2}^2(\Omega)+D_{0,-2}^2(\Omega)-D_{2,0}^2(\Omega)-D_{-2,0}^2(\Omega)\right)+\frac{1}{3}+\frac{1}{6}D_{0,0}^2(\Omega)\\
    \label{eq:cxz_cZz}
    \cos\theta_{Zz}\cos\theta_{Xz}&=&\cos\phi\sin\theta\cos\theta = \sqrt{\frac{1}{6}}\left(D_{-1,0}^2(\Omega)-D_{1,0}^2(\Omega)\right)\\
    \nonumber
    \cos\theta_{Zy}\cos\theta_{Xy}&=&-\sin\theta\sin\chi(\cos\phi\cos\theta\sin\chi+\sin\phi\cos\chi)=-\sqrt{\frac{1}{24}}\left(D_{-1,0}^2(\Omega)-D_{1,0}^2(\Omega)\right)+\\
    \label{eq:cxy_czy}
    &&\frac{1}{4}\left(-D_{-1,-2}^2(\Omega) -D_{-1,2}^2(\Omega)+D_{1,-2}^2(\Omega)+D_{1,2}^2(\Omega)\right).
  \end{eqnarray}
\end{widetext}
Therefore, the matrix elements of $\mathbf{H}_S$ and $\mathbf{H}_L(t)$ in the basis set of the eigenstates of the symmetric top basis can be computed using
\begin{eqnarray}
\nonumber
&&\int\mathrm{d}\Omega D_{M_1,K_1}^{J_1}(\Omega)D_{M_2,K_2}^{J_2}(\Omega)D_{M_3,K_3}^{J_3}(\Omega)=\\
\label{eq:three_wigner_integral}
&&8\pi^2\left(
\begin{array}{ccc}
J_1 & J_2 & J_3\\
K_1 & K_2 & K_3
\end{array}
\right)
\left(
\begin{array}{ccc}
J_1 & J_2 & J_3\\
M_1 & M_2 & M_3
\end{array}
\right),
\end{eqnarray}
where $\left(\begin{array}{ccc}
J_1 & J_2 & J_3\\
M_1 & M_2 & M_3
\end{array}\right)$ are the 3J symbols~\cite{Zare1988}.

\begin{acknowledgments}
The author acknowledges Dr. Johannes Flo{\ss}, Dr. Rosario Gonz\'alez-F\'erez and Prof. Dr. Lars Bojer Madsen for fruitful discussions and careful revision of the manuscript. This work was supported by NSERC Canada (via a grant to Prof. P. Brumer). The numerical results presented in this work were obtained at the Centre for Scientific Computing, Aarhus (Denmark).
\end{acknowledgments}

%

\end{document}